# Investigating Changes of Water Quality in Reservoirs based on Flood and Inflow Fluctuations

Shabnam Salehi*, Mojtaba Ardestani

School of Environment, College of Engineering, Isfahan University of Technology
* Corresponding Author: ssalehi1@crimson.ua.edu

**Abstract**

Water temperature and dissolved oxygen are essential indicators of water quality and ecosystem sustainability. Lately, heavy rainfalls are happening frequently and forcefully affecting the thermal structure and mixing layers in depth by sharply increasing the volume of inflow entitled "flash flood". It can occur by sudden intense precipitation and develop within minutes or hours. Because of heavy debris load and speedy water, this phenomenon has remarkable effects on water quality. A higher flow during floods may worsens water quality at lakes and reservoirs that are thermally stratified (with separate density layers) and decrease dissolved oxygen content. However, it is unclear how well these parameters represent the response of lakes to changes in volume discharge. To address this question, researchers simulate the thermal structure in two stratified reservoirs, considering the Rajae reservoir as a representative reservoir in the north of Iran and Minab reservoir in the south. In this study, the model realistically represented variations of dissolved oxygen and temperature of dams' Lake response to flash floods. The model performance was evaluated using observed data from stations on the dam's lake. In this case, the inflow charge considered in a 10-day flash flood from April 6th to April 16th during the yearly normal flow. The complete mixture in a part of the thermal structure has been proved in Rajaee reservoir. The nonpermanent impact of the massive inflow of storm runoff caused an increase in oxygen-consuming, leading to a severe decrease in dissolved oxygen on epilimnion and metalimnion. The situation in Minab reservoir was relatively different from Rajae reservoir. The inflow changes not only cause mixture but also help expanding stratification.

## 1. Introduction

Population growth is much more than available water resources in the world(Kummu et al. 2010). Reservoirs are artificial lakes, constructed to store water for future consumption. Altogether, reservoirs' operation has an essential role in water quality and quantity(Chaves, Tsukatani, and Kojiri 2004). For instance they can operate as trap for point source(PS) and non-point source(NPS) pollutant in highly agricultural basin that put them more sensitive to water quality changes (Wu and Liu 2012). Ivestigations also show, rivers can carry up to 90 percent of PS and NPS nutrient pollution loads into the lakes and reservoirs (Wu and Chen 2013) which put them more sensitive to water quality changes (Sohrabi et al,2022). Thermal stratification also influences the reservoir's quality of water and its ecology significantly (Gelda and Effler 2007; Chao et al. 2010).

The effect of temperature on water columns could change the chemical, biological, and physical characteristics of water (Lindim, Pinho, and Vieira 2011; Zouabi-Aloui, Adelana, and Gueddari 2015).

This change can be much more critical if the reservoir is stratified. Typically, water columns stratify during spring, summer, and autumn, including a surface layer (the epilimnion), with relatively warmer water, the bottom layer (the hypolimnion), which is colder, and a transient layer of water between the two, named "metalimnion"(Milstein and Zoran 2001; Saber, James, and Hayes 2018). At the same time, sufficient Dissolved Oxygen (DO) and prediction of DO concentrations are vital for aquatic managers responsible for maintaining a healthy ecosystem and water quality(Meding and Jackson 2003). Decomposition of organic materials and supporting the life of aquatic organisms are two of the crucial functions of dissolved oxygen, which makes it an unavoidable factor for maintaining high-quality water and a healthy ecosystem. Lower oxygen concentrations can cause the release of nutrients, and produce sulfide(Terry, Sadeghian, and Lindenschmidt 2017). Stronger stratification causes the increased stability of the water column, physically and confines the vertical exchange and mixing of water(Anohin et al. 2006; Wu et al. 2016). Typically, hypolimnetic DO, and its water quality are less as a result of a lower rate of mixing, water exchange, and darkness, which prevents photosynthesis (Elçi 2008; Hetherington et al. 2015; Milstein and Zoran 2001; Prats et al. 2018).

The thermal regime can be influenced by Many hydrological and meteorological parameters including air temperature and the temperature of inflow runoff, mode of reservoir operation, solar radiation and discharge(Butcher et al. 2015; Han et al. 2000; Wei, Dingguo, and Tao 2011; Saber, James, and Hayes 2018; Soleimani et al. 2016). The thermal regimes, which are also considered to be a key factor that affects thermocline thickness, stratification duration, and hydrodynamic circulation, are vulnerable to inflow discharge(Butcher et al. 2015; Han et al. 2000; Wei, Dingguo, and Tao 2011). High input which arises from runoff may severely disturb the stratification and promotes the mixing of water (Huang et al. 2014; Wang et al. 2012). Significant variations in water quality are problems associated with thermal stratification. Many of the factors not listed above, such as water chemistry, are themselves controlled in part by the temperature of the water. It is a serious problem for consumers who demand consistent high-quality water. There exist facilities designed to remove or modify thermal stratification thoroughly in some reservoirs. (Chapman and Organization 1996).

Different studies have examined water quality in a reservoir. Recently, the CE-QUAL-W2 model has used to examine selective withdrawal impacts on in-reservoir and water quality(Rheinheimer, Null, and Lund 2015; Zheng et al. 2017). Also, numerous studies have focused on the relationship between DO concentration, thermal stratification, and air temperatures; however, to our knowledge, this model has not been used to consider the effect of flash flood and elucidate river variation discharge impact directly on the temperature, and DO concentration. Intact aquatic ecology is a prerequisite to get benefits from a reservoir, so, understanding the effects of flash floods on water quality in the reservoirs are necessary to choose the optimal water intake position at the time of flash floods, to reduce the risk of water quality variations and water treatment costs.

The objective of this study is to determine the effect of inflow fluctuations on water quality of Rajae reservoir and validating results using Minab reservoir. In this paper, temperature and DO concentration employed as water quality parameters to compare with water inflow under flash flood situations.

## 2. Methodology

2.1. Overview

To simultaneously assessing the roles of factors, which influence dissolved oxygen in the reservoir, the CEQUAL-W2 model, combine different parameters such as inflow, reservoir geometry, water level, etc. It is a two-dimensional, laterally averaged, hydrodynamic and water-quality model that has been applied to reservoirs and lakes. Few previous studies examined the temperature and dissolved oxygen concentrations in a reservoir under flow fluctuation. While the CEQUAL-W2 model can simulate water conditions under flash flood, this is a relatively novel application of the model, and we used the model to develop a conceptual understanding of how oxygen and temperature changes occur in the Rajae reservoir as a major case study and Minab reservoir as a validator.

2.2. Study area

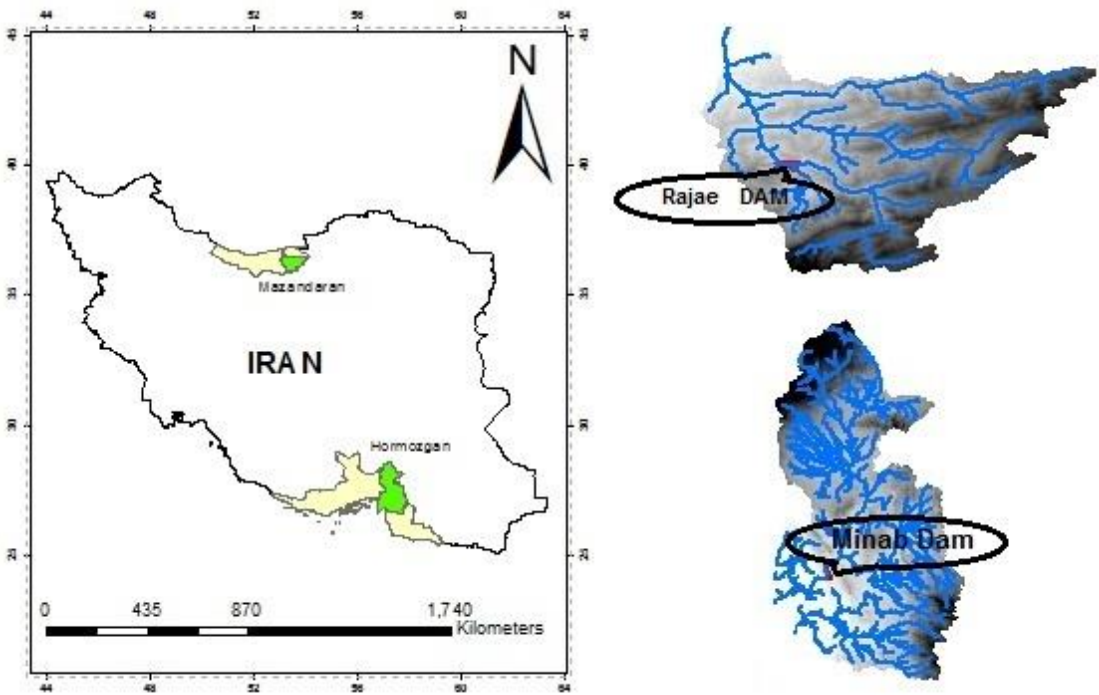

**Figure 1:** Rajaee and Minab reservoir location in Mazandaran and Hormozgan province.

Rajae dam (also called Soleyman Tangeh dam) is the most important source of drinking, industrial, and raw water for agricultural activities located in the south of Sari, Mazandaran province, with the geographical coordinate of (36° 14′ 55.82″ N, 53° 14′ 32.66″ E). It is of the highest importance in terms of flood control and power generation. It has a 5.2 Km$^2$ surface area and 133 meters height, with its full storage capacity near 165 Milion Cubic Meter (MCM). This dam is a warm monomictic stratified reservoir.

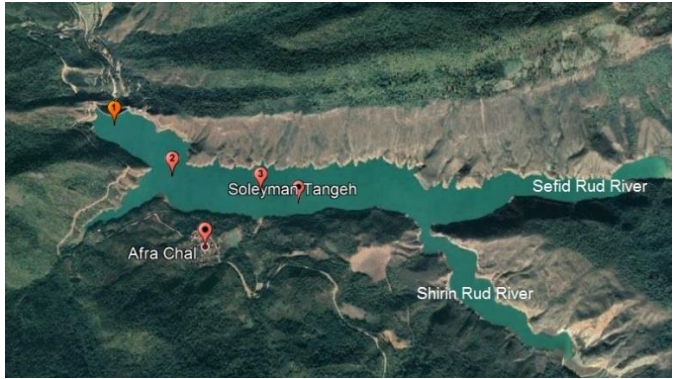

**Figure 2:** Map of Rajaee Reservoir and Rivers, data gathered from Afra Chal and Soleyman Tangeh stations, samples were obtained from No.1,2 and 3.

Shown in the Fig.2 is Tajan River, a permanent river born from the confluence of Sefid Rud River and Shirin Rud River, and is the foremost resource of Rajae dam. It finally drains into the Caspian Sea. This river originates from Alborz forested mountains and is 140 km long. It has a 4147 km$^2$ basin area and continues its course through agricultural lands. The mean water discharge of the dam is calculated to be 19.4 m$^3$s$^{-1}$ yearly (Saiidi et al. 2007), with plentiful variations. The maximum flow reaches the amount of 650.9 m$^3$s$^{-1}$, which was recorded once in 1979. The gathering location of meteorological and hydrological data has been shown in Fig.2. As seen in the figure, three sampling stations have been chosen for the calibrating the model. The geographic locations are described in Table.1.

**Table 1:** Location of sampling stations and data gathering

| Station Number | Latitude | Longitude |
| --- | --- | --- |
| S1 | 36°14'22.9776" | 053°15'55.0872" |
| S2 | 36°14'12.4764" | 053°16'08.1012" |
| S3 | 36°14'13.7904" | 053°16'09.7284" |
| Afra Chal | 36°14'12.3252" | 053°14'29.6808" |
| Soleiman Tangeh | 36°14'32.5860" | 053°15'03.6360" |

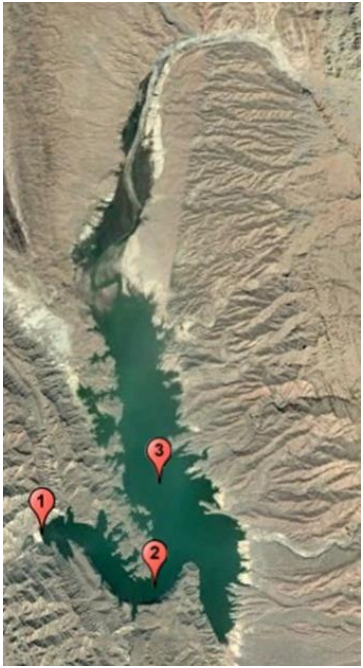

**Figure 3:** Map of Minab Reservoir and River, No.1, 2 and 3 are sampling locations

Minab dam (also named Esteghlal dam) is located in Hormozgan province, east of Minab town and south of Iran. The dam built in 1982 and functioned as a water producer for the drink and irrigation of Minab and Bandar Abbas cities. It also functions as a flood mitigator. With 57.19 $Km^2$ surface area and 59.25 meters height, its full capacity is 270MCM. The climate is warm and dry and characterized by high evaporation over precipitation. The river shown in Fig.3 is Minab River, which consists of two critical branches of Jaghin and Roudan. Three sampling stations have been selected to calibrate the model. The geographic locations describe in Table.2.

**Table 2:** Location of sampling stations

| Station Number | Latitude | Longitude |
|---|---|---|
| S1 | 27°9'50.57" | 57°6'45.95" |
| S2 | 27°9'26.30" | 57°7'57.65" |
| S3 | 27°10'24.27" | 57°7'51.37" |

2.3. Method

Three main steps have been followed in the research process. First, data gathering, second, modeling of the reservoir, and last the analysis of the data. Data were collected from 2001 to 2011 for Rajae reservoir and 2015 to 2019 for Minab reservoir. Parameters of water quality were provided by Mazandaran and Hormozgan Water Company for the period of simulation. Data regarding cloud cover, wind direction and speed, the dew and air point temperature, the quality of water entering the dam, and daily mean flow rates were applied in the modeling steps. To check the data accuracy, meteorology, flow, and water parameters reassumed to be considered as appropriate representations. For missing data, statistical relations were applied.

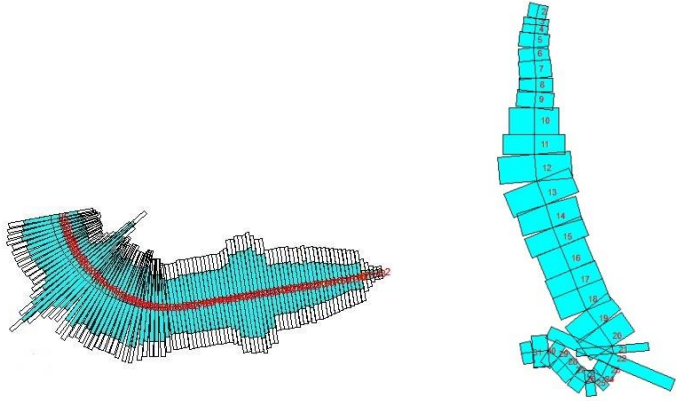

**Figure 4:** Plan view of segments in CE-QUAL-W2 model, Rajae reservoir (left) and Minab reservoir

Hydrodynamics, temperature, and DO were both simulated and calibrated according to the observed data achieved from sample stations, in order to confirm their accuracy. The hydrodynamic model was built and calibrated for the years between 2001 and 2011 for Rajaee reservoir and 2015 to 2019 for Minab reservoir. The bathymetric grid was generated using topographic maps in scale 1:100000. The water body was divided into 95 segments, each having a 50 meters length, and 45 layers, each with 1-meter thickness for Rajaee reservoir. The same was done on Minab dam, providing 32 segments and layers. The model length of segments in Minab reservoir varied between 100 to 600 meters. Area-depth volume curves, showing storage at different reservoir elevation, were used to check the bathymetry data.

2.4. Calibration

The calibration approach was used by trial and error to determine the appropriate model input , and to match the observed temperature and dissolved oxygen concentrations in the reservoir.

To be sure of model grid accuracy, its volume was compared to actual storage capacity. To simulate the thermal regime and DO concentrations, Mahab Qods Company measured the temperature and DO concentration profiles at both sites yearly at various depth intervals from top to bottom at three different monitoring stations and on multiple dates to compare the model outputs. This calibration was a complex process that used visual matching of the year-by-year variations in measurements and calculation of model. These efforts were aided by graphical profiles of temperature and DO, produced by model output that could be compared with the measured results [Figures 5 and 6].

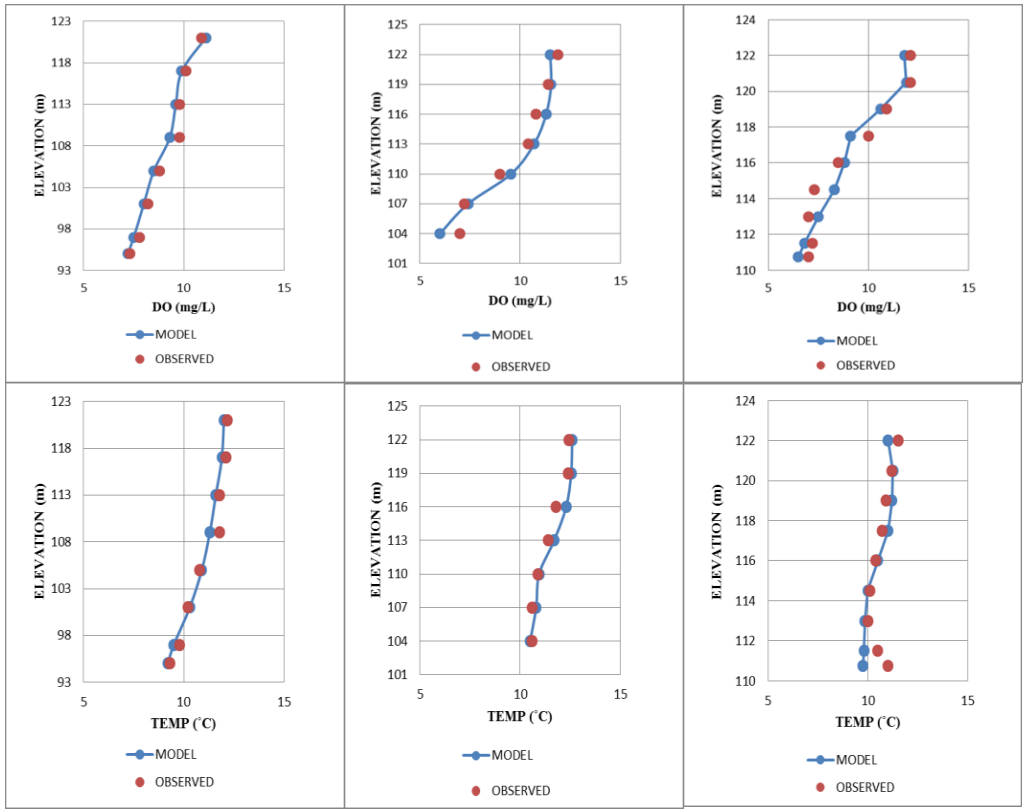

**Figure 5:** The compared modeled and observed temperature and DO; the pictures from left to right represent temperature and DO in three different stations (S1, S2, S3) in Rajaee reservoir

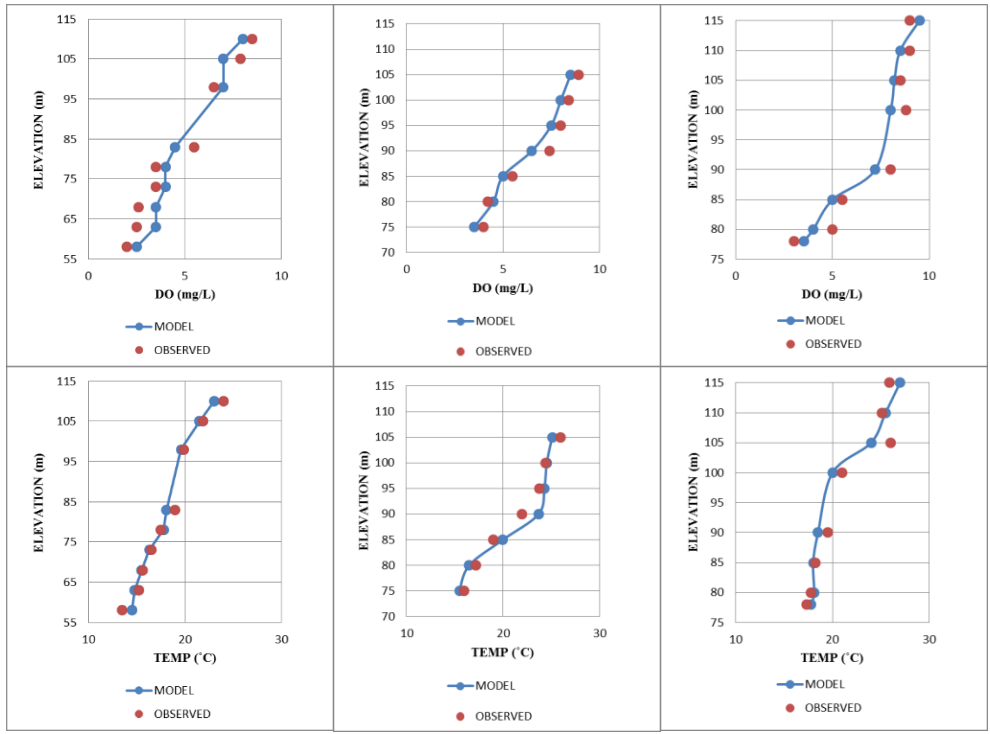

**Figure 6:** The compared modeled and observed temperature and DO; the pictures from left to right represent temperature and DO in three different stations (S1, S2, S3) in Minab reservoir

We have used the Absolute Mean Error (AME) and the Root Mean Square Error (RMSE) as statistical indices to compare the goodness-of-fit between our simulated and observed data. AME and RMSE were calculated according to the equations (1) and (2), and were used in the calibration process:

$$AME = \frac{\sum |predicted - observed|}{number\ of\ observations} \qquad (1)$$

## 2.5. Calibration validation

Model calibration can be defined as the modification of model parameters according to the comparative differences between observed and simulated data. Factors affecting the features of water temperature structure and water flow are various. A few to be mentioned are inflow, outflow, geometrical features, and meteorological conditions of the reservoir. The bathymetry, water surface, temperature, and dissolved oxygen are four parameters used for reservoir calibration. The calibration goal was to develop a simulation model robust enough to describe the temperature and DO concentration in the reservoirs. The temperature variations in the reservoir were related to the dissolved oxygen variations, showed by the calibration process. As a result, both temperature and dissolved oxygen were calibrated in the model, and the results regarding different parts of model calibration are as follows (Table.3).

**Table 3:** The results of Absolute Mean Error.

| Calibration Part (Rajaee/Minab)* | Absolute Mean Error (AME) |
|---|---|
| Bathymetry | 0.21/0.68 |
| Water Surface Layer | 0.66/0.11 |
| Temperature | 1.14/0.84 |
| Dissolved Oxygen | 0.97/1.03 |
| * Data provided between (2001-2011/2015-2019) | |

The calibration error range is entirely in the acceptable range (comparing to the models in the user manual of the CE-QUAL-W2)(Cole and Wells).

## 3. Results and discussion

3.1. Effects of flash flood on the temperature in Rajae reservoir

Fig.7 shows the temperature changes before, during, and after the flash flood, between April 6[th] and April 16[th]. During the stratification time, the inflow temperature is always less than the surface and higher than the bottom temperature. Firstly, the inflow enters the reservoir in the form of an underflow, then it forms an intrusional interflow, and finally, vertical mixing of the water column happens.

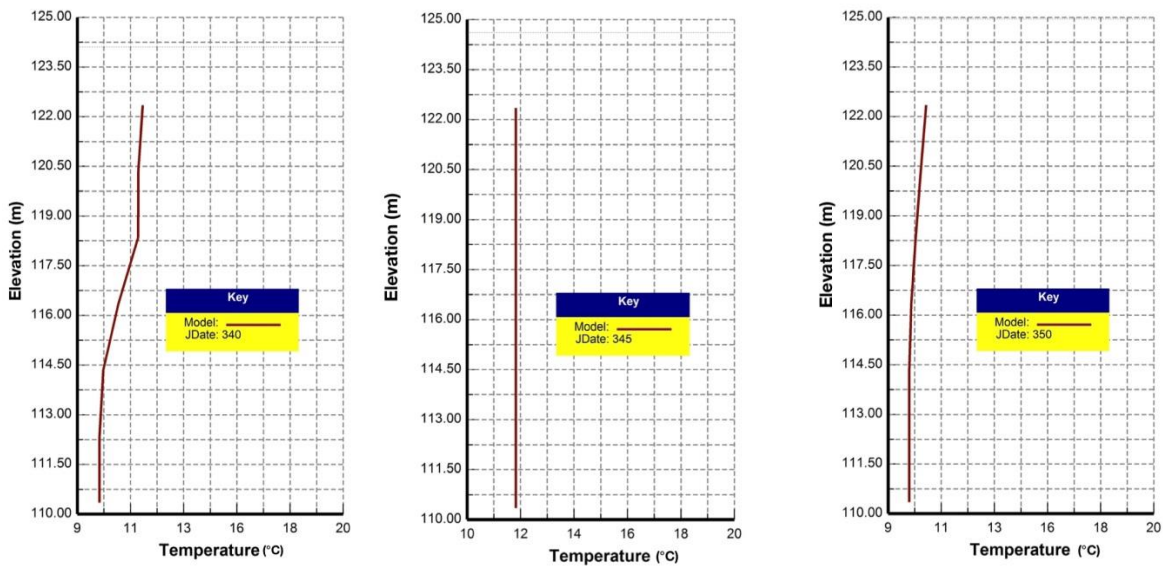

**Figure 7:** Temperature changes in depth before (day 340) , during (day 345) and after (day 350) flash flood at station 3.

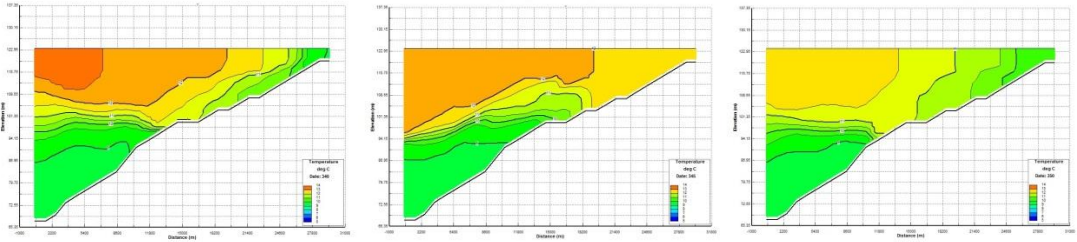

**Figure 8:** Comparison of DO profiles with respect to distance along the reservoir from the dam over the reservoir's depth in days 340, 345 and 350

Fig.8 portray the temperature changes under inflow variations. Minimum changes observe near the dam between 0 to 92 m sea level. Flood impact appears from 12000 m and it causes to mixing layer from the water surface to 20 m in depth. These pictures show how temperature layers mix and shape new stratification and decrease the average temperature in part of the reservoir.

3.2. Effect of flash flood on dissolved oxygen in Rajae reservoir

The typical thermal stratification confines the oxygen transfer between the deep layer (hypolimnion) and the surface layer. Organic materials and other oxygen-consuming components in the water and sediment gradually decrease the DO content of hypolimnion. Overall, the release of pollutants can directly affect and change the hypolimnion's available DO, given the change from 8.75 mg/L on April 6th to a lower measure of 4.75 mg/L on April 16th in 2011. Because of continuous stormwater runoff in April 2011, the hypolimnion zone was damaged, and in the time following the flash flood, DO of the bottom water remained in the range of 7-9 mg/L.

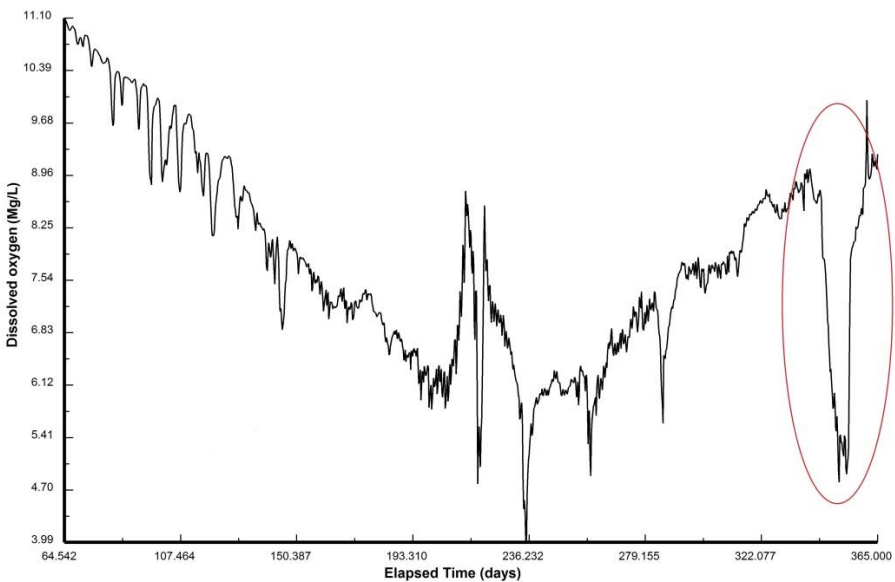

**Figure 9:** Dissolved oxygen during the simulation's time; the oval shows DO concentration during flash flood time.

The trend of DO changes during dry and wet seasons are shown in year 2011 in the reservoir [Figure 8]. Before precipitation increases, DO concentration decreased from 11.1 mg/L to 6.12 mg/L because of the lower volume of inflow. The inflow starts to increase during the wet season that is followed by precipitations. After the flash flood, in the last days of March, a decrease in water DO from 9.3 mg/L to 5 mg/L was observed. Five days after the second flash flood, the DO content decreased again [Figure 9]. The water runoff causes organic materials to enter the reservoir, increasing the oxygen use-up suddenly.

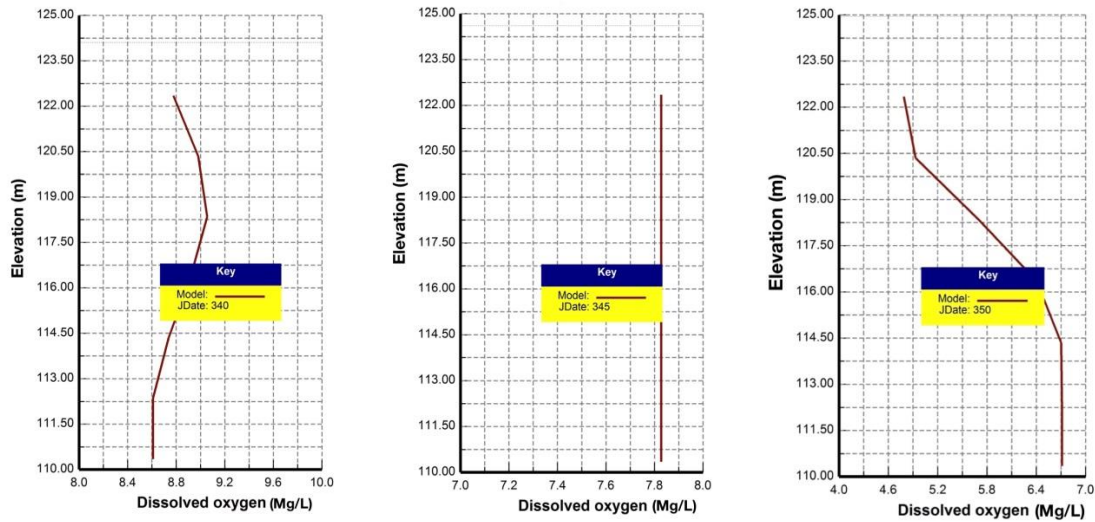

**Figure 10:** Dissolved oxygen changes in depth before (day 340), during (day 345) and after (day 350) flash flood at station 3.

To better understand the effect of inflow fluctuations on DO concentrations, Fig.11. shows DO of Rajaee's reservoir on the 6[th] and 16[th] of April with average water level. The oxygen stratification at various elevations and distances from the dam is visible. There is pronounced variation of DO have seen from the bottom of the reservoir, and the thickness of this layer is almost 15m at a distance between 8600 m. The oxygen concentration varied between less than 1 to 3 mg/L . This area located between 95 and 110 m above sea level.

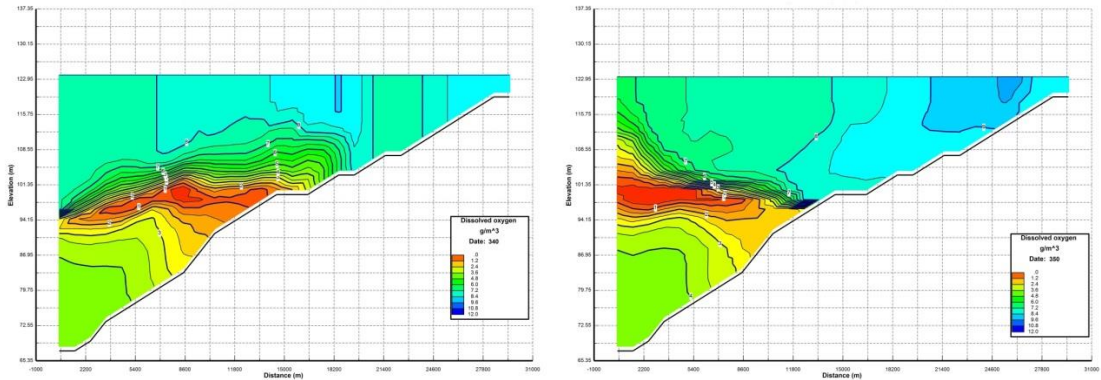

**Figure 11:** Dissolved oxygen in reservoir (Left) April 6, before flash flood and (Right) April 16, after flash flood

Fig.11(right) shows the same condition except for the arrival of the flash flood and the high inflow rate, which changed the stratification of the reservoir. The anoxic area started at zero meter distance from the dam and continued until around 8600m distance. The thickness of this zone is comparatively reduced (from 93 m to 103 m elevate, the thickness is around 10 m), demonstrating the mixture of this layer with more oxygenated lower and higher levels and the extension of it to the nearer distances to the dam body. The more deep layers remained intact with oxygen levels between 4 to 6 mg/L. Because of the mixture between deoxygenated water and oxygenated superficial layers, the DO of superficial layers has been reduced.

Oxygen transfer has been limited from the surface to deeper water by stratification. The DO concentration near the bottom of the reservoir at distance 9500 to 13000 m was lower than 1 mg/L and became anoxic. During the flash flood, the stratified zone was damaged in a way that the anoxic zone starts to move and mixes with other layers.

3.3. Effect of flash flood on dissolved oxygen in Minab reservoir

The same situation has been applied in the Minab reservoir to examine and compare results. Fig.12 shows the level of DO concentrations in a different part of the reservoir in a normal situation. The stratification is quite apparent.

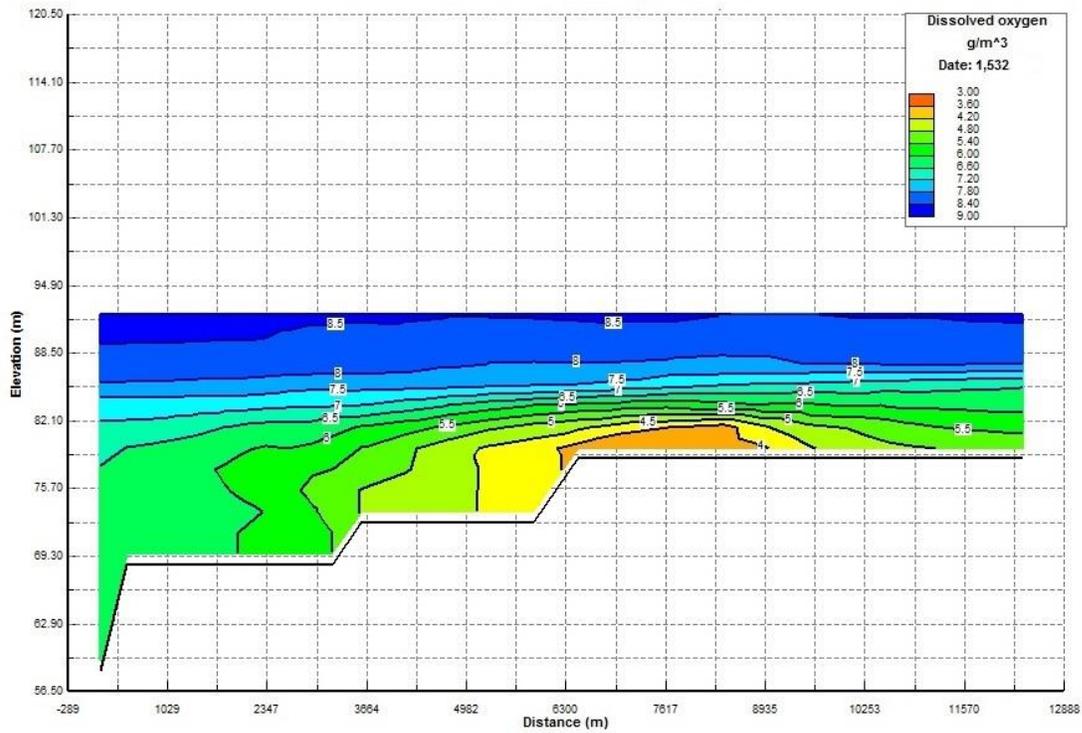

**Figure 12:** Dissolve oxygen in the reservoir.

For understanding the effect of flash flood, researchers simulate artificial flash flood to find out the changes of DO concentration and stratification response to flood. Fig.13 shows DO concentration after flash flood on the same day.

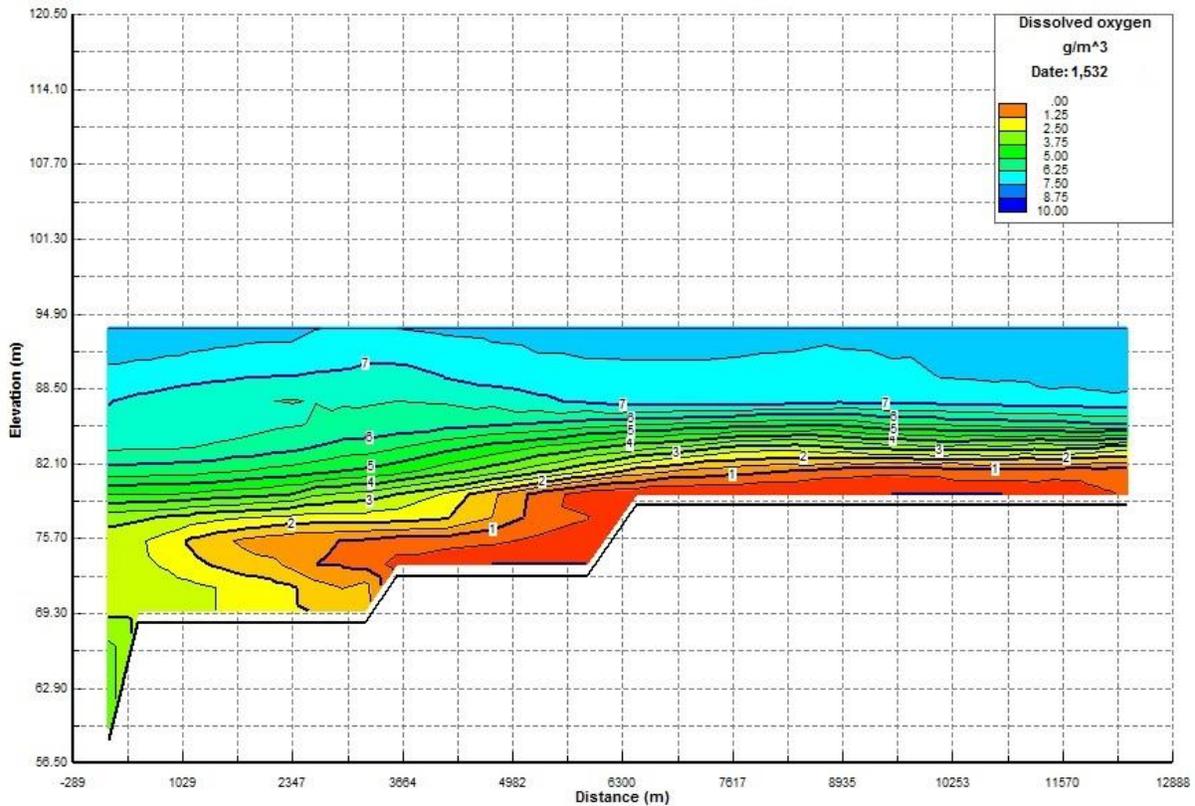

**Figure 13:** Dissolve oxygen in Minab reservoir under the flood situation.

The DO concentration was between 4 to 9.5 mg/L before the flash flood and decreased to a level between 0 and 7 mg/L after the flash flood. The transit layers (metalimnion) become condenser comparing to hypolimnion and epilimnion. Overall, the deeper layers were more affected and deoxygenated under flood situation. Despite the common saying that flood always helps the mixturing, it has been shown that the flood caused expanding of stratification and decreased the quality of water in the hypolimnion layer of Minab reservoir.

3.4. Effect of flash flood on Temperature in Minab reservoir

Fig.14 show the contour of temperature in a different part of the modeled reservoirs before and after a flood. There were no observed changes, except that flood caused an expansion in stratification and flood water intruded in deeper layers.

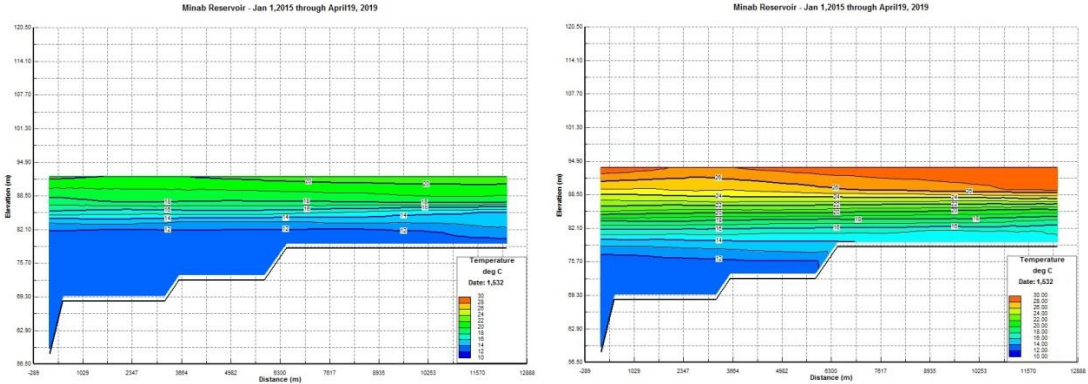

**Figure 14:** Temperature contours in Minab reservoir before (left) and after flood

## 4. Conclusion

High inflow volumes can significantly reduce the stability of stratification and induce early mixing in Rajae reservoir because of water flow can intensify vertical disturbance. Herein, it has been shown that the lower the initial water level, the initial stratification of the reservoir is more distinguishable, and the effect of the flood on the stratification is stronger. Sohrabi et al. (2023) demonstrate the impact of tropical cyclones on various water bodies in their paper. Therefore, as a result of intense inflow volume fluctuations, due to heavy precipitation, the thermal regime of Rajae's reservoir has become much more sensitive to change.

Heavy rainfall events associated with massive erosion and high velocity of the water can cause increasing nutrient input into the reservoir and increases the temperature. Increasing temperature will decrease the solubility of oxygen and other gases (Perlman 2013). The water will not hold enough oxygen for aquatic organisms to survive if it is too warm. Higher water temperatures, in addition to its effects on aquatic organisms, can increase the solubility and thus toxicity of certain compounds such as lead, cadmium, and zinc as well as compounds like ammonia (Cooke et al. 2016).

The temperature of water influences the organisms' tolerance limit. In higher temperatures, mortality rates are significantly increased. This occurs because of oxygen consumption, tissue permeability, and metabolic rate all increase with increased water temperature (Bhadja and Vaghela 2013).

Because of massive erosion from upstream of the river, flash flood events import a very high nutrient input into the reservoir mainly in particulate form. Decreasing density flow is an acceptable option to reduce the input of particulate associated pollutants. However, the effectiveness and conditions needed to be provided are still unclear and need further assessments. As discussed earlier, the high inflow during flash floods affects the water stratification and the quality of water in each layer. The CE-QUAL-W2 model is able to simulate the temperature and DO concentration in both reservoirs in good agreement with the field observations. The results of this study demonstrate the higher potential of changes in the hypolimnion in both reservoir and uncertainty about stratification response to inflow fluctuations.

The model was used to understand oxygen concentrations under the inflow changes because both reservoir has had quality problems with rapid increase inflow. The model allows us to investigate and compare the importance of individual factors on temperature and dissolve oxygen . It is important to note that the model simplifies the many and complex processes that occur in the reservoir.

Attempts to model the reservoir with regard to the DO and temperature, during heavy inflows can give further insight into choosing the appropriate intake height and reaching the high-quality water. The selective intake height can improve reservoir outflow DO concentration up to 30% and the general quality of outflow up to 77% depends on operational strategies(Aalami, Abbasi, and Nourani 2018). This is an influential step in decreasing the treatment costs to a lower optimum.

**Conflict of interest**: The authors declare no conflict of interest